\documentclass[12pt,preprint]{aastex}

\def\axp{\hbox{AX~J1844.8$-$0256}}
\def\taxp{\hbox{XTE~J1810$-$197}}

\newcommand\xte{{\it RXTE\/}}
\newcommand\chandra{{\it Chandra}}
\newcommand\xmm{{\it XMM-Newton}}

\shorttitle{Radio Emission from TAXP XTE J1810$-$197}
\shortauthors{Halpern et al.}

\slugcomment{To appear in The Astrophysical Journal Letters}
\begin{document}

\title{Discovery of Radio Emission from Transient Anomalous
X-ray Pulsar XTE J1810$-$197} 

\author{J. P. Halpern\altaffilmark{1}, E. V. Gotthelf\altaffilmark{2},
R. H. Becker\altaffilmark{3}, D. J. Helfand\altaffilmark{1}, and
R. L. White\altaffilmark{4}}

\altaffiltext{1}{Department of Astronomy, Columbia
University, 550 West 120th Street, New York, NY 10027;
jules,djh@astro.columbia.edu}
\altaffiltext{2}{Columbia Astrophysics Laboratory, Columbia University,
550 West 120th Street, New York, NY 10027;
eric@astro.columbia.edu}
\altaffiltext{3}{Physics Department, University of California,
1 Shields Avenue, Davis, CA 95616; and Institute of Geophysics and
Planetary Physics, Lawrence Livermore National Laboratory, Livermore,
CA 94550; bob@igpp.ucllnl.org}
\altaffiltext{4}{Space Telescope Science Institute, 3700 San Martin Drive, 
Baltimore, MD 21218; rlw@stsci.edu}

\begin{abstract}
We report the first detection of radio emission from any anomalous
X-ray pulsar (AXP).  Data from the Very Large Array (VLA) MAGPIS survey
with angular resolution $6^{\prime\prime}$ reveals a point-source
of flux density $4.5 \pm 0.5$ mJy at 1.4 GHz at the precise location 
of the 5.54~s pulsar \taxp.
This is greater than upper limits from
all other AXPs and from quiescent states of soft gamma-ray repeaters (SGRs).
The detection was made in 2004 January, 1~year
after the discovery of \taxp\ during its only known outburst.
Additional VLA observations both before and after the outburst yield
only upper limits that are comparable to or larger than the
single detection, neither supporting nor ruling out a decaying radio
afterglow related to the X-ray turn-on.  Another hypothesis is that,
unlike the other AXPs and SGRs, \taxp\ may power a radio synchrotron
nebula by the interaction of its particle wind with a moderately
dense environment that was not evacuated by previous activity from this 
least luminous, in X-rays, of the known magnetars.
\end{abstract}

\keywords{pulsars: general --- pulsars: individual (XTE J1810$-$197,
AX J1844.8$-$0256)  --- radio continuum: stars --- X-rays: stars}

\section{Introduction}

According to the magnetar theory of AXPs and SGRs \citep{dun92},
the high-energy radiation of these slow
pulsars ($5 < P < 12$\,s) is supplied not by magnetic braking of
their rotation, which is insufficient, but from particle acceleration
and/or internal heating that is powered by the decay of an enormous
magnetic field ($B \geq B_{\rm QED} = 4.4 \times 10^{13}$~G). 
The absence of radio emission is a defining characteristic of AXPs.
Radio emission from magnetars has only been seen following impulsive
high-energy events from two SGRs.  Specifically, a radio source
detected within a few days of the intense 1998~August~27 $\gamma$-ray burst
from SGR~1900+14 faded rapidly during the following month \citep*{fra99}.
The giant flare of SGR~1806$-$20 on 2004~December~27
\citep{hur05,pal05} was accompanied by a radio source that exhibited
a complex decay \citep{cam05,gae05b}.  No radio transient
has been recorded from an AXP, e.g., after a bursting episode of 1E~2259+586
\citep{kas02}, and no {\it persistent\/} radio emission, whether
compact or extended, has been clearly attributed to an AXP or an SGR.

The transient AXP \taxp\ is a 5.54~s X-ray pulsar discovered
by \cite{ibr04} using the {\it Rossi X-ray Timing Explorer} (\xte),
and located to subarcsecond precision by \chandra\ and infrared 
counterpart detections \citep{got04b,isr04}.  When discovered in a high state, 
the source was already decaying exponentially with a time constant of 
$270$~days \citep{ibr04} from a recorded maximum flux $F_X (2-10\ {\rm keV})
\approx 6 \times 10^{-11}$ ergs~cm$^{-2}$~s$^{-1}$.  The onset
and initial properties of the outburst were not observed,
but \xte\ scans restrict the epoch of the event to between 2002 November
and 2003 January.  Archival X-ray detections of the previously anonymous 
source indicate a steady
quiescent state with flux two orders-of-magnitude less than the maximum
measured value.  Detailed interpretation of the decaying X-ray outburst of
\taxp\ using \xmm, as well as its historical X-ray properties, were reported
by \citet{got04b}, \citet{hal05}, and \citet{got05}.

This Letter presents a unique VLA detection of \taxp\ 1~year
after its X-ray outburst, and upper limits from other archival VLA data
that restrict its spectrum and pattern of variability.
Possible origins of the radio emission are explored,
and comparisons are made to upper limits from other AXPs, and to transient
radio detections that followed outbursts of SGRs.  Although the 
distance to \taxp\ is highly uncertain, we adopt here $d=2.5$~kpc following 
the argument in \citet{got05}.  This is a revision of the larger estimates
used in earlier papers.

\section{Observations and Analysis}

\begin{figure}
\plotone{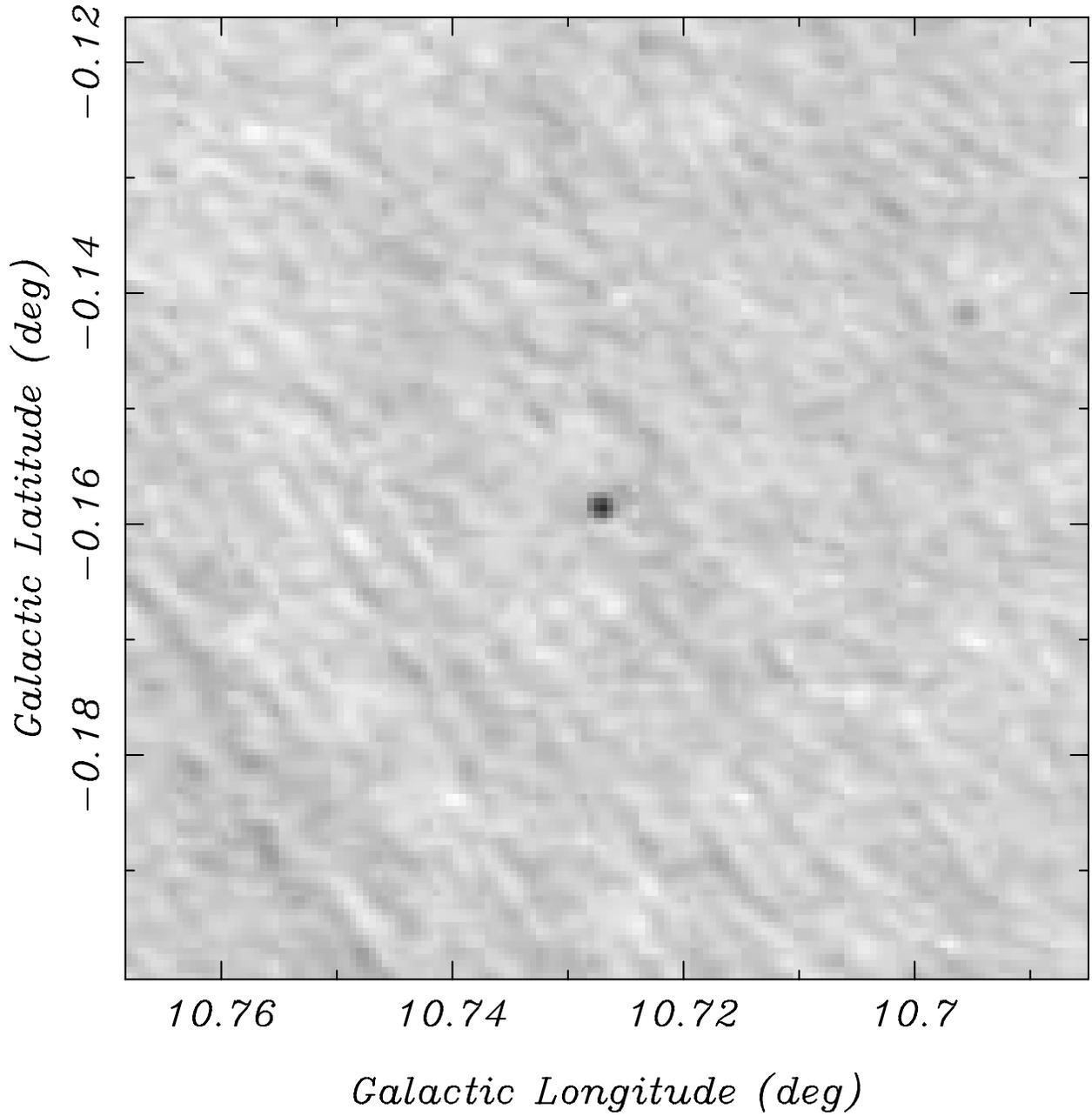}
\caption{MAGPIS VLA image at 1.4 GHz in B configuration of
\taxp\ taken in 2004 January.  A $9 \sigma$ radio
source with flux density $4.5\pm0.5$~mJy is present
within $1^{\prime\prime}\!.0$ of the X-ray/IR coordinates.
This source is unresolved by the
$6^{\prime\prime}$ FWHM beam.}
\label{image}
\end{figure}

Observations at 1.4~GHz of the region containing \taxp\ 
were obtained on several days in 2004 January as part of
the VLA Multi-Array Galactic Plane Imaging Survey
\citep*[MAGPIS;][]{hel05}\footnote{http://third.ucllnl.org/gps}
A total of 15 min of exposure were
accumulated in B configuration, with a resulting beam
${\rm FWHM} = 6^{\prime\prime}$ and rms sensitivity 0.5~mJy per beam. 
Figure~\ref{image} shows the resulting image centered on the
\chandra\ coordinates of \taxp\ given in \citet{got04b},
R.A. $18^{\rm h}09^{\rm m}51.\!^{\rm s}08$, Decl.
$-19^{\circ}43^{\prime}51.\!^{\prime\prime}7$ (J2000.0).
A definite radio source is detected with flux density $4.5 \pm 0.5$ mJy
at coordinates R.A. $18^{\rm h}09^{\rm m}51.\!^{\rm s}14$, Decl.
$-19^{\circ}43^{\prime}51.\!^{\prime\prime}2$ (J2000.0),
within $1^{\prime\prime}$ of the \chandra\ position, which
in turn is identical to the coordinates of its IR counterpart \citep{isr04}.
The difference is comparable to the combined astrometric
uncertainty of both instruments.
Since the radio source is the only one within a $3^{\prime}\!.5$
radius of the pulsar, the probability that it is a chance coincidence
of an unrelated object within a radius of $1^{\prime\prime}$
can be estimated at $\sim 2 \times 10^{-5}$, and the identification
on the basis of position alone is compelling.
The radio source extent is unresolved, from which we infer an
upper limit on its intrinsic radius of $3^{\prime\prime}$, or 
$r < 1.1 \times 10^{17}\,d_{2.5}$~cm, where $d_{2.5}$ is the
distance in units of 2.5~kpc.

We then analyzed several VLA observations from earlier Galactic
plane surveys \citep*{whi05} and other archival data,
with results listed in Table~\ref{obslog}.  Only upper
limits are obtained from these observations, including
the ones reported by \citet{gae03}.
When quoted at the $3 \sigma$ level, nearly all of the upper
limits are higher than the single detection.
Therefore, they contain little information about intrinsic variability
of the source.  However, two VLA upper limits obtained in 2003,
following the X-ray outburst of \taxp, restrict
an hypothesized power-law decay to $t^{-0.6}$ or slower.
The upper limit of 3.6~mJy in 2004 March is then
in mild conflict with the detection in 2004 January.
If, on the other hand, the source is truly 
constant, then the spectrum from $0.33-4.8$~GHz cannot be fitted with 
a single power law of the form $F_{\nu} \propto \nu^{\alpha}$.
The upper limit of 1~mJy at 4.8~GHz requires $\alpha < -1.1$ between
1.4 and 4.8~GHz, while the upper limit of 10.5~mJy at 330~MHz
requires $\alpha > -0.66$ between 330~MHz and 1.4~GHz.  The implied
spectral curvature might be intrinsic, or the onset of
self absorption; alternatively, variability is required.

We also inspected the MAGPIS 1.4~GHz B configuration image for another 
transient AXP candidate, \axp\ \citep{got98,tor98},
which yields a $3 \sigma$ upper limit of 0.5~mJy for a point source
at R.A. $18^{\rm h}44^{\rm m}54.\!^{\rm s}69$, Decl.
$-02^{\circ}56^{\prime}53.\!^{\prime\prime}4$ (J2000.0).
We derived this position from the sum of several archival \chandra\ images.
This 6.97~s pulsar associated with a supernova remnant \citep*{gae99}
remains only a candidate AXP because its period derivative has not yet
been measured due to a dramatic reduction in its X-ray flux \citep*{vas00}.

\section{Discussion}

No radio pulsations have been detected from AXPs,
with pulsed flux density limits $< 0.1$~mJy at 1.4~GHz
obtained for most sources \citep{mer02}.
It is not yet established by statistics alone whether the small set of
AXPs are intrinsically radio quiet or just unfavorably oriented,
their long periods implying small active polar caps and narrow beams.
Neither has persistent radio emission
been seen from this class until now, although flux limits
from interferometric imaging, listed in Table~\ref{compare},
are generally not as deep as the single-dish searches for pulsed flux.
Although the VLA source associated with \taxp\ is spatially unresolved,
the bulk of its flux is unlikely to be due to magnetospheric
pulsar emission.  A 4.5~mJy source pulsed at the 5.54~s X-ray
period should be easily detected in the Parkes Multibeam Pulsar Survey
operating at 1.4~GHz, as have other long-period, high $B$-field
pulsars \citep[e.g.,][]{mcl03} that are at least an order-of-magnitude
fainter.  Therefore, we assume that most of the emission is
incoherent synchrotron radiation arising from well outside the
light cylinder, $r_{\rm lc} = 2.6 \times 10^{10}$~cm, but within
the VLA resolution limit $r = 10^{17}$~cm.  Its brightness temperature 
is then $T_{\rm B} = 100\,(d_{2.5}/r_{17})^2$~K, and it is probably larger
than $10^{12}$~cm, assuming that $T_{\rm B} < 10^{12}$~K.

To put the unique radio detection of \taxp\ in context, we note
that its radio flux density is larger than all of the
upper limits from other AXPs (Table~\ref{compare}),
but its radio luminosity is demonstrably larger than only two or three,
considering that the
distances to AXPs, as tabulated by \citet{mer02} and \citet{woo06},
may be uncertain by a factor of 2.  The radio detection
of \taxp\ may be a significant clue into the physics of its class. 
\taxp\ differs from other AXPs in two respects.  First, it is transient,
with an X-ray luminosity that is comparable to the persistent AXPs only 
near the peak of its outburst.
Second, its quiescent state may be long-lived, so that its time-averaged
X-ray luminosity could be two orders-of-magnitude less than the others.
The only other transient X-ray source in Table~\ref{compare} is \axp,
but its distance is uncertain and possibly large \citep{gae99}, so its
radio luminosity is not necessarily smaller than that of \taxp.

An obvious hypothesis, that the 2004 January radio luminosity
of \taxp\ is a transient afterglow that originated with the 2003
January X-ray outburst, is problematic.
First, there is the twice documented absence of a more luminous radio
source in the months following the X-ray outburst, in 2003 February and
April.  It is also notable that no radio source, to a limit of
$50\,\mu$Jy, was detected 2 days after an X-ray outburst and glitch of
the AXP 1E~2259+586 \citep{kas02}.
Second, there was no evidence for any soft $\gamma$-ray outburst
from \taxp\ resembling the flares of the SGRs, only a slowly decaying,
thermal X-ray spectrum that integrated over a time scale of 300 days
to a total energy of $\sim 4 \times 10^{42}\,d^2_{2.5}$ ergs \citep{got05}.
In fact, it is possible that \taxp\ was {\it never} more luminous than
a normal AXP.  Its activity is perhaps better characterized as a 
turn-on than an outburst, and is in marked contrast to the $10^{46}$
ergs $\gamma$-ray flare of SGR 1806$-$20 in 2004 December \citep{hur05,pal05}.

Third, only in the aftermath of
giant flares from the SGR 1806$-$20 and 1900+14 have decaying
radio sources been detected by the VLA, and even these were
faint and short-lived.  In the case of SGR 1900+14,
the peak 1.4~GHz flux density 10 days after the 1998 August flare 
was 0.7~mJy \citep{fra99}, but it decayed as a power law of index
$-2.6$ and was no longer detectable after 1~month.  Just 1.6 days
after another burst activation from SGR 1900+14 in 2001 April,
no radio emission was detected to a $5 \sigma$ limit of 0.45 mJy at 5~GHz
\citep{kou01}, which was not unexpected since the $\gamma$-ray
fluence of the 2001 burst was 25 times smaller than the 1998 burst.
The 2004 December giant flare from SGR 1806$-$20 produced
a radio source initially detected at 178 mJy at 1.4~GHz,
which faded to 10 mJy after 50 days \citep{cam05}.
Extrapolated to 1~year, SGR 1806$-$20 should be as faint
in the radio as \taxp\ was
1~year after its X-ray turn-on.  This despite the fact that
the $\gamma$-ray flare of SGR 1806$-$20
was three orders-of-magnitude more luminous
than the entire extrapolated X-ray luminosity of \taxp\ from
the peak of its turn-on to the present time.

For the sake of completeness, we note that at
least the kinematic properties of the SGR events
are not incompatible with the size of the \taxp\ radio source.
If we assume that the radio emitting plasma observed in 2004 January
was first ejected from the neutron star a year earlier,
then we can place an upper limit on the ejection velocity,
$v/c < 0.12\,d_{2.5}$.   This is comparable to the velocity directly
estimated in the case of SGR 1806$-$20 from its resolved VLA images,
$v/c = (0.27\pm0.10)\,d_{15}$ \citep{gae05b}.

If the radio emission from \taxp\ is not a transient afterglow,
it may be caused by a variable wind of
high-energy particles, which we hypothesize could be responsible
for powering a small synchrotron emitting nebula under favorable
circumstances.  The spin-down rates of AXPs fluctuate by a large factor,
which is taken as evidence that
rotational energy is regularly removed by mechanisms other than dipole
braking, e.g., a particle wind \citep*{tho98,har99}.
Continued monitoring by \xte\ and \xmm\ shows that the
spin-down rate of \taxp\ varies irregularly by at least a factor of 4,
with $\dot P = (0.5-2.2) \times 10^{-11}$~s~s$^{-1}$ 
\citep{ibr04,got04b,got05}.  Assuming a moment of inertia $I = 10^{45}$
g~cm$^2$, this implies the possibility
of a wind of luminosity $\dot E = 4\pi^2 I \dot P/P^3 \sim 4 \times 10^{33}$
ergs~s$^{-1}$, which is comparable to the quiescent thermal X-ray 
luminosity \citep{hal05}, and five orders-of-magnitude
larger than the radio luminosity $\nu L_{\nu} \approx 5 \times 10^{28}$
ergs~s$^{-1}$ at 1.4~GHz.
In addition, \xte\ continued to find small X-ray bursts
from \taxp\ long after its initial outburst.
\citet{woo05} document four such bursts in 2003 September, 2004 February,
and 2004 May, from a total exposure time of 9 days.
Although these have thermal spectra and blackbody radii much smaller
than the surface area of a neutron star, their temperatures of 4--8~keV
imply local flux greater than the Eddington value.
Whether their initial trigger is magnetic reconnection or crust fracturing,
they likely cause matter to be ejected from the surface.
The strongest of these bursts had a energy
of $3.5 \times 10^{37}\,d_{\rm 2.5}^2$ ergs, and together they
emitted $4.2 \times 10^{37}\,d_{\rm 2.5}^2$ ergs.  Therefore
we can estimate a time-averaged luminosity in bursts of 
$\dot E \sim 5.4 \times 10^{31}$ ergs~s$^{-1}$, which is two
orders-of-magnitude less than the estimated wind luminosity.

It is possible that long-lived radio emission is
powered by the accumulated plasma ejected in the process
of creating the X-ray bursts, as well as by the more luminous wind.
Following the arguments of \citet{fra97} and \citet{gae00},
the conversion of the ubiquitous pulsar winds to an observable
pulsar wind nebula (PWN)
may be limited not by the energy carried in the wind,
but by the environment -- specifically, by whether the
ambient pressure is large enough to shock the wind and confine the
nebula.  In the case of \taxp, balancing the assumed wind power
$\dot E \sim 4 \times 10^{33}$ ergs~s$^{-1}$ at a termination
shock radius $r_{\rm w} < 10^{17}$~cm requires a pressure 
$\dot E/4\pi r_{\rm w}^2c = 1.1 \times 10^{-12}\,(r_{\rm w}/10^{17})^{-2}$ dynes~cm$^{-2}$.
This is a rather modest value even if the true radius of the nebula
is only $10^{16}$~cm and the pressure is $1 \times 10^{-10}$ dynes~cm$^{-2}$,
or alternatively, if the wind power has been 100 times larger than assumed here.
In comparison, \citet*{hel01} showed that the radius of the Vela PWN
is consistent with the thermal pressure in the surrounding hot
supernova remnant, $8.5 \times 10^{-10}$ dynes~cm$^{-2}$
measured from X-ray spectroscopy \citep{mar97}.  There
is no evidence for a supernova remnant surrounding \taxp, which
may imply that the surrounding temperature is cooler.  If we assume
an ionized medium of $T = 10^4$~K, then the required hydrogen density
is $n_{\rm H} \sim 0.4 (r_{\rm w}/10^{17})^{-2} (T/10^4)^{-1}$ cm$^{-3}$.
The actual development of a PWN is a time-dependent
process, with a forward shock radius estimated as
$r_{\rm s}(t) = (\dot E/4\pi\rho_0)^{1/5}t^{3/5}$ \citep{aro83}.
For the numerical values assumed here, $r_{\rm s} \leq 10^{17}$~cm 
is reached in $\leq 100$~years.
Alternatively, if the pulsar's space velocity $v_{\rm p}$ exceeds 
$\dot r_{\rm s}$, then a bow-shock nebula will form with apex radius 
$r_{\rm a} = (\dot E/4\pi\rho_0 v_{\rm p}^2)^{1/2}$ that is smaller
than $r_{\rm s}$ for a ``static'' nebula.

We note that the radio emitting efficiency of \taxp, as defined by
\citet{fra97} and \citet{gae00}, is comparable to the upper limits
obtained by them for rotation-powered pulsars of similar $\dot E$
as we are assuming for the wind of \taxp. Therefore, evidence is
lacking that the average wind power of \taxp\
is much greater than its spin-down value.  Additionally, there
is no X-ray PWN around \taxp, whose image
is consistent with a point-source at the \chandra\ resolution of
FWHM $\approx 0^{\prime\prime}\!.6$ \citep{isr04}.
Only pulsars with $\dot E > 10^{36}$ ergs~s$^{-1}$
have X-ray PWNe \citep{got04a}.
Perhaps the difference between \taxp\ and other magnetars is that
\taxp\ is surrounded by a dense medium, while the past
activity of the other, more energetic magnetars, has disrupted and
evacuated their surroundings. We note that this is different
from the hypothesis advanced by \citet{gae05a} to explain an apparent
\ion{H}{1} cavity around AXP 1E~1048.1$-$5937, which they supposed was
blown by the wind of its massive progenitor star.

\section{Conclusions and Future Work}

The detection of radio emission from an AXP
comes as a surprise because it is contrary to
a defining property of its class.  But the physical
basis of the classification of \taxp\ is
unchanged.  The long period, unsteady spin-down,
X-ray luminosity exceeding its spin-down power, and detailed
properties of its two-component X-ray spectrum, are all squarely in 
the AXP mold, and support the magnetar interpretation \citep{got05}.
Although the radio luminosity of \taxp\ is energetically insignificant,
a more detailed investigation of its origin could lead to a better 
understanding of the evolution of AXPs and their immediate surroundings.  

The spectral and spatial structure of this compact radio source can
be studied with multifrequency observations in the VLA A configuration,
which has the potential to resolve the source if it is a PWN powered
by a wind of comparable luminosity to the X-ray luminosity
or the spin-down power.  If size and spectral information are obtained,
the particle and magnetic field energies of this presumed synchrotron
source can be determined.
While the evidence for temporal variability is marginal,
variability would not be unexpected, as the X-ray source is still declining
from its 2003 January turn-on.   A monitoring program can study the
relationship, if any, of the radio structure and luminosity to the
X-ray history.  However, it is possible that the environment
and the normal X-ray quiescence of \taxp, rather than its unusual outburst,
is what enables it to support a radio source.

Having suggested that \taxp\
is the prototype of a major subclass of young pulsars that
may have been detected but not yet recognized in quiescence \citep{got04b},
we now anticipate the prospect of an additional marker, 
in the form of a compact radio source, that will aid in their discovery.

\acknowledgements

This work was supported by NASA ADP grant NNG05GC43G to JPH and EVG.
RHB and DJH acknowledge the support of the National Science Foundation
under grants AST-02-6-309 and AST-02-6-55, respectively.

\clearpage 

\begin{deluxetable}{lccccc}
\tablecolumns{6} 
\small
\tablewidth{0pt}
\tablecaption{VLA Observations of \taxp}
\tablehead{
\colhead{Dates}  &  \colhead{Frequency}  &  \colhead{Array}   &  \colhead{Beam FWHM}   &  \colhead{$F_{\nu}$} & \colhead{References} \\
\colhead{}      &  \colhead{(GHz)}      &  \colhead{Configuration}
&  \colhead{($^{\prime\prime}$)}        &  \colhead{(mJy)} 
}
\startdata
1983 Dec      & 1.42   & B    & 6            & $< 5.5$     & 1,2,3 \\
1990 Oct      & 4.82   & CnB  &  6            & $< 1$       & 1,3,4  \\ 
2001 Jan -- 2002 Aug   & 0.33 & A,B,C,D & 25 & $< 10.5$    & 5,6 \\
2003 Feb  & 1.42   & DnC  & $62\times35$ & $< 10.5$    & 5,6 \\
2003 Apr & 1.42   & D    &   45           & $< 7$       & 1,7   \\
2004 Jan  & 1.42   & B    & 6            & $4.5\pm0.5$ & 1,7   \\
2004 Mar      & 1.42   & C     &   15           & $< 3.6$     & 1,7 
\enddata
\tablerefs{(1) This work; (2) Zoonematkermani et al. 1990;
(3) White et al. 2005; (4) Becker et al. 1994; (5) Gaensler \& Brogan 2003;
(6) Brogan et al. 2004; (7) Helfand et al. 2005.}
\tablecomments{Upper limits quoted here are $3\sigma$ based on
the rms noise per beam.  Uncertainty quoted on the detection is $1\sigma$.} 
\label{obslog}
\end{deluxetable}

\clearpage 

\begin{deluxetable}{lcccc}
\tablecolumns{5} 
\small
\tablewidth{0pt}
\tablecaption{Imaging Observations of AXPs at 1.4 GHz}
\tablehead{
\colhead{Source} &  \colhead{$F_{\nu}$} & \colhead{$d$} &
\colhead{$F_{\nu} d^2$} & \colhead{References} \\
\colhead{} & \colhead{(mJy)} & \colhead{(kpc)} & \colhead{(mJy kpc$^2$)}
}
\startdata
4U 0142+61              & $< 0.16$      & 3   & $< 1.4$  & 1 \\
1E 1048.1$-$5937        &  ...          & 2.7 & ...  & 2 \\
1RXS J170849.0$-$400910 & $< 1.8$       & 5   & $< 45$  & 1 \\
XTE J1810$-$197\tablenotemark{a}         & $4.5 \pm 0.5$ & 2.5 & 28 & 3 \\
1E 1841$-$045           & $< 0.36$      & 7   & $< 18$ & 4 \\
AX J1844.8$-$0256\tablenotemark{a}       & $< 0.5$  & $\sim 8$ & $< 32$ & 3  \\
1E 2259+586             & $< 0.050$     & 3   & $< 0.45$  & 5,6
\enddata
\tablenotetext{a}{Transient X-ray source.}
\tablerefs{(1) Gaensler et al. 2001; (2) Gaensler et al. 2005a;
(3) This work; (4) Kriss et al. 1985; (5) Coe et al. 1994;
(6) Kaspi et al. 2002.}
\tablecomments{Upper limits quoted here are 
for unpulsed emission at the $3\sigma$ level.} 
\label{compare}
\end{deluxetable}

\clearpage 

\clearpage 

\begin{thebibliography}{}

\bibitem[Arons(1983)]{aro83}Arons, J. 1983, \nat, 302, 301

\bibitem[Becker et al.(1994)]{bec94}
Becker, R. H., White, R. L., Helfand, D. J., \& Zoonematkermani, S.
1994, \apjs, 91, 347

\bibitem[Brogan(2004)]{bro04}Brogan, C. L., Devine, K. E.,
Lazio, T. J., Kassim, N. E., Tam, C. R., Brisken, W. F.,
Dyer, K. K., \& Roberts, M. S. E. 2004, \aj, 127, 355

\bibitem[Cameron et al.(2005)]{cam05}Cameron, P. B., et al.
2005, \nat, 434, 1112

\bibitem[Coe et al.(1994)Coe, Jones, \& Lehto]{coe94}
Coe, M. J., Jones, L. R., \& Lehto, H. 1994, \mnras, 270, 178

\bibitem[Duncan \& Thompson(1992)]{dun92}Duncan, R. C., \&
Thompson, C. 1992, \apjl, 392, L9

\bibitem[Frail et al.(1999)Frail, Kulkarni, \& Bloom]{fra99}
Frail, D. A., Kulkarni, S. R., \& Bloom, J. S. 1999, \nat, 398, 127

\bibitem[Frail \& Scharringhausen(1997)]{fra97}Frail, D. A., \&
Scharringhausen, B. R. 1997, \apj, 480, 364

\bibitem[Gaensler \& Brogan(2003)]{gae03}Gaensler, B. M.,
\& Brogan, C. L. 2003, Astron. Telegram, 207, 1

\bibitem[Gaensler et al.(1999)Gaensler, Gotthelf, \& Vasisht]
{gae99}Gaensler, B. M., Gotthelf, E. V., \& Vasisht, G. 1999,
\apj, 526, L37

\bibitem[Gaensler et al.(2005a)]{gae05a}Gaensler, B. M.,
McClure-Griffiths, N. M., Oey, M. S., Haverkorn, M., Dickey,
J. M., \& Green, A. J. 2005a, \apjl, 620, L95

\bibitem[Gaensler et al.(2001)]{gae01}Gaensler, B. M., Slane,
P. O., Gotthelf, E. V., \& Vasisht, G. 2001, ApJ, 559, 963

\bibitem[Gaensler et al.(2000)]{gae00}Gaensler, B. M., Stappers, W.,
Frail, D. A., Moffett, D. A., \& Johnston, S. 2000, \mnras, 318, 58

\bibitem[Gaensler et al.(2005b)]{gae05b}Gaensler, B. M., et al. 2005b,
\nat, 434, 1104

\bibitem[Gotthelf(2004)]{got04a}Gotthelf, E. V. 2004,
in IAU Symp. 218, Young Neutron Stars and Their Environments,
ed. F. Camilo \& B. M. Gaensler (San Francisco: ASP), 225 

\bibitem[Gotthelf \& Halpern(2005)]{got05}Gotthelf, E. V., \&
Halpern, J. P. 2005, \apj, 632, 000 (astro-ph/0506511)

\bibitem[Gotthelf et al.(2004)]{got04b}Gotthelf, E. V., Halpern, J. P.,
Buxton, M., \& Bailyn, C. 2004, \apj, 605, 368

\bibitem[Gotthelf \& Vasisht(1998)]{got98}Gotthelf, E. V., \&
Vasisht, G. 1998, NewA, 3, 293

\bibitem[Halpern \& Gotthelf(2005)]{hal05}Halpern, J. P., \&
Gotthelf, E. V. 2005, \apj, 618, 874

\bibitem[Harding et al.(1999)Harding, Contopoulos, \& Kazanas]{har99}
Harding, A. K., Contopoulos, I., \& Kazanas, D. 1999, \apjl, 525, L125

\bibitem[Helfand et al.(2005)Helfand, Becker, \& White]{hel05}
Helfand, D. J., Becker, R. H., \& White, R. L. 2005, in preparation

\bibitem[Helfand et al.(2001)Helfand, Gotthelf, \& Halpern]{hel01}
Helfand, D. J., Gotthelf, E. V., \& Halpern, J. P. 2001, \apj, 556, 380

\bibitem[Hurley et al.(2005)]{hur05}Hurley, K., et al. 2005, \nat,
434, 1098

\bibitem[Ibrahim et al.(2004)]{ibr04} Ibrahim, A. I., et al. 2004, 
\apjl, 609, L21

\bibitem[Israel et al.(2004)]{isr04}Israel, G. L., et al. 2004,
\apjl, 603, L97

\bibitem[Kaspi et al.(2002)]{kas02}Kaspi, V. M., Gavriil, F. P.,
Woods, P. M., Jensen, J. B., Roberts, M. S. E., \& Chakrabarty, D.
2002, \apjl, 588, L93

\bibitem[Kouveliotou et al.(2001)]{kou01}Kouveliotou, C., Tennant, A.,
Woods, P. M., Weisskopf, M. C., Hurley, K., Fender, R. P.,
Garrington, S. T., Patel, S. K., \& G\"o\u g\"u\c s, E. 2001, \apjl, 558, L47

\bibitem[Kriss et al.(1985)]{kri85}Kriss, G. A., Becker, R. H.,
Helfand, D. J., \& Canizares, C. R. 1985, \apj, 288, 703 

\bibitem[Markwardt \& \"Ogelman(1997)]{mar97}Markwardt, C. B., \&
\"Ogelman, H. B. 1997, \apj, 480, L13

\bibitem[McLaughlin et al.(2003)]{mcl03}McLaughlin, M. A., et al. 2003,
\apjl, 591, L135

\bibitem[Mereghetti et al.(2002)]{mer02}Mereghetti, S., Chiarlone, L.,
Israel, G. L., \& Stella, L. 2002, in Proc. 270th WE-Heraeus Seminar
on Neutron Stars, Pulsars, and Supernova Remnants, ed. W. Becker,
H. Lesch, \& J. Tr\"umper (MPE Rep. 278; Garching: MPE), 29

\bibitem[Palmer et al.(2005)]{pal05}Palmer, D. M., et al. 2005, \nat,
434, 1107

\bibitem[Thompson \& Blaes(1998)]{tho98} Thompson, C., \& Blaes, O.
1998, \prd, 57, 3219

\bibitem[Torii et al.(1998)]{tor98}Torii, K., Kinugasa, K.,
Katayama, K., Tsunemi, H., \& Yamauchi, S. 1998, \apj, 503, 843

\bibitem[Vasisht et al.(2000)]{vas00} Vasisht, G., Gotthelf, E. V.,
Torii, K., \& Gaensler, B. M. 2000, \apj, 542, L49

\bibitem[White et al.(2005)White, Becker, \& Helfand]{whi05}
White, R. L., Becker, R. H., \& Helfand, D. J. 2005, \aj, 130, 586

\bibitem[Woods et al.(2005)]{woo05}Woods, P. M., Kouveliotou, C.,
Gavriil, F. P., Kaspi, V. M., Roberts, M. S. E., Ibrahim, A.,
Markwardt, C. B., Swank, J. H., \& Finger, M. H. 2005, \apj, 629, 985

\bibitem[Woods \& Thompson(2006)]{woo06}Woods, P. M., \& Thompson, C.
2006, in Compact Stellar X-ray Sources, ed. W. H. G. Lewin \& M. van der
Klis (Cambridge: Cambridge Univ. Press), in press (astro-ph/0406133)

\bibitem[Zoonematkermani et al.(1990)]{zoo90}Zoonematkermani, S.,
Helfand, D. J., Becker, R. H., White, R. L., \& Perley, R. A. 1990,
\apjs, 74, 181

\end{thebibliography}
\end{document}